\documentclass{article}
\usepackage{spconf}
\usepackage{amssymb,amsmath,bm}
\usepackage{xcolor}
\usepackage{multirow,siunitx,etoolbox}
\usepackage{graphicx}
\usepackage{float}
\usepackage{csquotes}
\usepackage{booktabs}

\title{An Analysis of Speech Enhancement and Recognition Losses in Limited Resources Multi-talker Single Channel Audio-Visual ASR }
\name{Luca Pasa$^{1,3}$, Giovanni Morrone$^2$, Leonardo Badino$^1$}
\address{
  $^1$Istituto Italiano di Tecnologia, Ferrara, Italy\\
  $^2$Department of Engineering "Enzo Ferrari", University of Modena and Reggio Emilia, Modena, Italy\\
  $^3$Department of Mathematics, University of Padova, Italy}

\begin{document}
%
\maketitle

\begin{abstract}
In this paper, we analyzed how audio-visual speech enhancement can help to perform the ASR task in a cocktail party scenario. Therefore we considered two simple end-to-end LSTM-based models that perform single-channel audio-visual speech enhancement and phone recognition respectively. Then, we studied how the two models interact, and how to train them jointly affects the final result. 

We analyzed different training strategies that reveal some interesting and unexpected behaviors. The experiments show that during optimization of the ASR task the speech enhancement capability of the model significantly decreases and vice-versa. Nevertheless the joint optimization of the two tasks shows a remarkable drop of the Phone Error Rate (PER) compared to the audio-visual baseline models trained only to perform phone recognition. We analyzed the behaviors of the proposed models by using two limited-size datasets, and in particular we used the mixed-speech versions of GRID and TCD-TIMIT.

\end{abstract}


\section{Introduction}

Although state-of-the-art speech recognition systems have reached very high accuracy, their performance  drops significantly when the signal is recorded in challenging conditions (e.g. mismatched noises, low SNR, reverberation, multiple voices). On the other hand, humans show a remarkable ability in recognizing speech in such conditions (\textit{cocktail party effect} \cite{mcdermott}).

Some robust ASR systems process the audio signal through a speech enhancement or separation stage before passing it to the recognizer \cite{narayanan2014investigation}. An alternative approach is to train the ASR model in a multi-task fashion where speech enhancement/separation and recognition modules are concatenated and jointly trained \cite{wang2015joint, narayanan2015improving, chen2015speech}. 


Several recent works showed significant advancements in speech separation \cite{DANet17, Isik2016SingleChannelMS, Kolbaek17, luo2018tasnet} and target speaker extraction \cite{zmolikova2017speaker, Wang2019} from mixed-speech mixtures. 

These works proposed end-to-end models and training strategies that are exploited to perform multi-speaker \cite{qian2018single, seki2018} and target speaker speech recognition \cite{delcroix2018single}.

The aim of the paper is to study how the speech enhancement task can help in recognizing the phonetic transcription of the utterance spoken by target speaker from single-channel audio of several people talking simultaneously. Note that this is an ill-posed problem in that many different hypotheses about what the target speaker says are consistent with the mixture signal. We addressed this problem by exploiting the visual information associated to the speaker of interest in order to extract her speech from input mixed-speech signal.
In \cite{morrone2019face} we demonstrated that face landmark's movements are very effective visual features for the enhancement task when the size of the training dataset is limited. 

In the last few years many audio-visual approaches have shown remarkable results by using neural networks to solve speech-related tasks with different modalities of the speech signal. These include audio-visual speech recognition \cite{chung2017lip, afouras2018deep}, audio-visual speech enhancement \cite{gabbay2018visual, michelsanti2019training, ochiai2019multimodal} and audio-visual speech separation \cite{ephrat_looking_2018, afouras_conversation:_2018, owens2018audio}.

It is well know that simultaneously learning multiple related tasks from data  can be more advantageous rather than learning these tasks independently  \cite{evgeniou2004regularized}. The class of these methods belong to Multi-Task Learning (MTL)  \cite{zhang2017survey}. 

Several speech processing applications are tightly related, so MTL methods can improve performance and reduce generalization error. In particular, robust ASR models show better accuracy when they are trained with other tasks \cite{wang2015joint, chen2015speech, tang2016multi}.

An MTL LSTM-based model is proposed in \cite{chen2015speech}, where the cost function is the weighted sum of ASR and speech enhancement losses. 
Some of these methods differ from the most common MTL approaches, where the differentiation of tasks is made only in the last layers of the network. These methods are also referred to as \enquote{joint learning}. 

We study how the speech enhancement and recognition tasks interact using an approach that belongs to this class of methods. The approach is equivalent to merging two different models with different loss functions: one to optimize the speech enhancement, and one for the phone recognition task. Our aim is to analyze the interaction between the ASR and enhancement tasks, and understand whether (and how) it is advantageous to train them jointly. For this reason, we firstly train and analyze a simple ASR model, then we study whether adding a preliminary speech enhancement stage helps in performing the ASR task. In order to analyze how the two tasks (and the respective loss functions) interact we propose three different training techniques that allow to unveil the strengths and the weaknesses of this approach. In particular we focused our attention on a very common audio-visual setting  where the quantity of available data for training the model is limited.

\setlength{\abovedisplayskip}{1pt}
\setlength{\belowdisplayskip}{1pt}
\vspace{-1.5em}
\section{Models Architecture}
\vspace{-0.5em}
In this section we present the models used to analyze and study how speech enhancement and recognition tasks can be combined to perform phone recognition in a cocktail party scenario. In order to perform a fair analysis, we use very simple and common architectures based on deep Bi-directional Long Short-Term Memory (BLSTM) \cite{graves2005framewise}. 
These models are fed with the sequence $\mathbf{s}=[ \mathbf{s}_1, \dots , \mathbf{s}_T ]$ where  $\mathbf{s}_i \in \mathbb{R}^N, \forall i \in [1,\dots,T]$ and/or the sequence ${\textbf{v} = [\textbf{v}_1, \dots ,\textbf{v}_T],} \; {\textbf{v}_t \in \mathbb{R}^M}$. $\mathbf{s}$ represents a spectrogram of the mixed-speech audio input, $T$ is the number of frames of the spectrogram and $\textbf{v}$ is the motion vector computed from the video face landmarks \cite{Kazemi_2014_CVPR} of the speaker of interest.

\subsection{ASR Model}
The ASR model consists of a deep-BLSTM. It first computes the mel-scale filter bank representation derived from the spectrogram $\mathbf{s}_i$:
\begin{equation}
    \mathbf{s}^m_i = \mathbf{m} \cdot \mathbf{s}_i, \label{eq:mel}
\end{equation}
where $\mathbf{m} \in \mathbb{R}^{C \times N}$ is the matrix that warps the spectrogram to the mel-filter banks representation.

We developed 3 different versions of this model that differ by the input used to perform the ASR task. The first version only uses acoustic features, therefore  $\mathbf{x}_i^{asr}=\mathbf{s}^m_i$.

The second version uses both audio and visual  features, thus:
$\mathbf{x}_i^{asr} = \begin{bmatrix}
   \mathbf{s}^m_i \\
   \mathbf{v}_i \\
 \end{bmatrix}, \; \; \mathbf{x}_i^{asr}  \in \mathbb{R}^{C+M}.\nonumber
$

The last version of the ASR models only fed with motions vector computed from face landmarks: $\mathbf{x}_i^{asr}=\mathbf{v}_i$.

All the models map $\mathbf{x}_i^{asr}$ to the phone label $\hat{\mathbf{l}}_i$ by using $Z^{asr}$ BLSTM layers. The output of the last BLSTM layer is linearly projected onto $\mathbb{R}^P$ in order to use the CTC loss.
This ASR model can be defined as follows: \mbox{$\; \mathcal{F}^{asr}( \mathbf{x}_i^{asr} , \theta^{asr}) = \hat{\mathbf{l}}_j$}. Where $\theta^{asr}$ is the set of parameters of the ASR model.
The model uses a CTC loss function to optimize the phone recognition task: $\mathcal{L}^{asr}(\mathbf{l}_j,\hat{\mathbf{l}}_j) =  {CTC_{loss}}(\mathbf{l}_j,\hat{\mathbf{l}}_j)$.

\vspace{-0.5em}
\subsection{Enhancement Model}
The Enhancement model is developed with the goal of denoising the speech of the speaker of interest given the mixed-speech input. The model input at time step $i$ is:
\mbox{$\mathbf{x}_i = \begin{bmatrix}
   \mathbf{s}_i \\
   \mathbf{v}_i \\
 \end{bmatrix}, \; \; \mathbf{x}_i  \in \mathbb{R}^{N+M}.
$}
The speech enhancement task target is $\mathbf{y} = [ \mathbf{y}_1, \dots \mathbf{y}_T ],$ where $\; \mathbf{y}_i \in \mathbb{R}^N$ is a slice of the spectrogram of the clean utterance spoken by the speaker of interest.
The enhancement model consists of $Z^{enh}$ BLSTM layers and a final layer that projects the output onto $\mathbb{R}^N$. This last layer uses sigmoid as activation function and, in order to obtain values in a scale comparable to the speech enhancement target, it multiplies the output by $k \cdot \mathbf{d}$, where $k$ is a constant and $\mathbf{d} \in \mathbb{R}^N$ is a vector that contains the standard deviations of each output feature. 
The enhancement model can be defined as a function: $\sigma(\mathcal{F}^{enh}(\mathbf{x}_i, \theta^{enh})) \odot (k \cdot \mathbf{d}) = \hat{\mathbf{y}}_i$, 
where $\sigma$ is the sigmoidal function and $\theta^{enh}$ is the set of parameters of the model. As loss function the model uses the Mean Squared Error (\textit{MSE}): $\mathcal{L}^{enh}(\mathbf{y}_i,\hat{\mathbf{y}}_i) =  {MSE}(\mathbf{y}_i,\hat{\mathbf{y}}_i).$\\
\vspace{-2em}
\subsection{Joint Model}
In order to evaluate whether and how speech enhancement can help in performing ASR in cocktail party scenario, we developed a model that is the combination of the Enhancement model and the ASR model: 
$\mathcal{F}^{asr}( \mathbf{m} \cdot \hat{\mathbf{y}}_i , \theta^{asr}) = \hat{\mathbf{l}_j}$. 
Note that only the enhancement part of the model exploits the visual information, while the ASR part receives in input only the output of the audio enhancement module $\hat{\mathbf{y}}_i$.

\vspace{-1em}
\subsection{Training Strategies}
Our aim is to explore and study the behaviors of the two losses $\mathcal{L}^{enh}$ and $\mathcal{L}^{asr}$. Therefore, we explored different techniques to perform training in order to analyze how the two losses interact.

The first training technique, henceforth referred to as  \textit{joint loss}, consists of using a loss that is a weighted sum of the two loss functions, 
$\mathcal{L}_{join} = \lambda \cdot \mathcal{L}^{enh} + \mathcal{L}^{asr}$,

where $\lambda \in \mathbb{R}$ is the coefficient that multiplies $\mathcal{L}^{enh}$.

During training we observed that the ratio of the two losses significantly changes.
To keep both the two losses at the same level of magnitude we also experimented with an adaptive coefficient 
\begin{equation}
    \lambda_{adapt}= 10^{\lfloor \log_{10}(\mathcal{L}^{asr})\rfloor} / 10^{\lfloor \log_{10}(\mathcal{L}^{enh}) \rfloor}. \label{eq:lambda}
\end{equation}

The second training method, \textit{alternated training}, consists of alternation of the speech enhancement and ASR training phases. This training procedure performs a few steps of each phase several times. The speech enhancement phase will use $\mathcal{L}^{enh}$ as loss function and  therefore only $\theta^{enh}$ parameters will be updated during this phase. During the ASR phase the loss function will be $\mathcal{L}^{asr}$. 
A particular case of the \textit{alternated training} is the \textit{alternated two full phases training} where the two phases are performed only one time each for a large number of epochs.

In \textit{alternated training} and \textit{alternated two full phases training}, the $\mathcal{L}^{asr}$ optimization phase updates both $\theta^{enh}$ and $\theta^{asr}$ parameters. 
For both techniques we also developed a \textit{weight freezing} version that optimize $\mathcal{L}^{asr}$ by only updating $\theta^{asr}$.

\section{Experimental Setup}
In this section, we report and discuss all the results obtained during the analysis.

\subsection{Dataset}
We decided to focus our analysis on a challenging and common scenario where the quantity of available data and resources is limited. Indeed, we performed the analysis by using the GRID \cite{cooke_audio-visual_2006} and TCD-TIMIT \cite{harte_tcd-timit:_2015} audio-visual limited-size datasets. We used the mixed-speech speaker-independent versions of these two datasets proposed in \cite{morrone2019face} as a starting point and then added the phone transcriptions for the speaker of interest. The GRID and TCD-TIMIT dataset were respectively split into disjoint sets of $25$/$4$/$4$ and $51$/$4$/$4$ speakers for training/validation/testing respectively.

For both datasets we used standard TIMIT phone dictionary. In particular in GRID the number of used phones is limited to 33 (as the vocabulary is limited to few tens of words), while in TCD-TIMIT all the 61 TIMIT phones are present.
Similarly to what is usually done with TIMIT, the 61 phones were mapped to 39 phones after decoding, when computing the Phone Error Rate (PER).




\subsection{Baseline and Model Setup}
In order to create a strong baseline to evaluate the performance of the joint model, we tested the various versions of the ASR model.
All these baseline models consist of 2 layers of 250 hidden units and were trained by using back-propagation through time (BPTT) with Adam optimizer. 
For what concerns the joint model, we used the same number of layers for both ASR and enhancement components: $Z^{enh} = Z^{asr} = 2$. Each layer consists of 250 hidden units with $\tanh$ activation function. We performed a limited random search-based hyper-parameter tuning, therefore all reported results may be slightly improved.

\subsection{Phone Error Rate Evaluation}

\setlength\tabcolsep{6pt}
\begin{table}
  \centering
  \begin{tabular}{l|S|S @{\hspace{0.7\tabcolsep}} S}
    \toprule
    \multirow{2}{*}{} &
      \multicolumn{1}{c|}{GRID} &
      \multicolumn{2}{c}{TCD-TIMIT} \\
    {Training Method} & {PER} & {PER-61} & {PER-39}\\
    \midrule
    ASR-Mod. Clean-Audio & 5.8 &  46.7	& 40.6 \\ 
    ASR-Mod. Mixed-Audio & 49.4 & 78.4 & 71.3\\ 
    ASR-Mod. Mixed-A/V & 49.9  & 77.2 & 70.9\\ 
    ASR-Mod. Visual & 29.4 & 78.6 & 74.7\\
    \midrule
    Joint-Mod. Joint loss & 15.4  & 53.1 & 47.7 \\ 
    Joint-Mod. Alt. 2 full & 16.0  & 45.6 & 41.2\\ 
    Joint-Mod. Alt. 2 full freeze & 18.7  & \hspace{0.3em}$\mathbf{44.3}$ & \hspace{0.3em}$\mathbf{40.0}$\\ 
    Joint-Mod. Alt. & \hspace{-0.05em}$\mathbf{13.9}$  & 44.9 & 40.6\\ 
    Joint-Mod. Alt. freeze & 18.1  & 61.3 & 55.5\\ 
    Joint-Mod. PIT Alt. & 43.3  & 67.1 & 62.4\\ 
    \bottomrule
  \end{tabular}
  \caption{Results on GRID and TCD-TIMIT, the first part of the table contains the results by the ASR baseline models, while in the second part, the results obtained by the joint models trained with the various training strategies are reported. All the results are computed on the test set.}
  \label{tab:Res_per}
  \vspace{-2mm}
\end{table}
\setlength{\belowcaptionskip}{-10pt}

Table~\ref{tab:Res_per} reports PER of all baseline models and of the joint models with different training strategies.
Note that the results on GRID obtained by using visual input can not be compared with the results obtained in \cite{assael2016lipnet} since our model was trained with a significantly smaller version of the dataset.

It is also important to point out that in the ASR-Model fed with Mixed-Audio/Video input the visual information does not help to reach better results, while in \cite{morrone2019face} we show that the visual information is very effective in performing speech enhancement.

The joint model achieved on TCD-TIMIT a PER that is comparable with the clean-audio baseline, while results on GRID are slightly worse but still much better than baseline results. 
Note that the difference in the achieved PER between the two datasets is mainly due to the difference of vocabulary size (GRID has a tiny 52 word vocabulary), phonotactics (as in GRID the word sequences are more constrained) and utterance lengths, indeed, the length of the sequences is variable in TCD-TIMIT, while it is fixed in GRID.   

\begin{figure}[ht]
\centering
\includegraphics[width=2.8in,keepaspectratio]{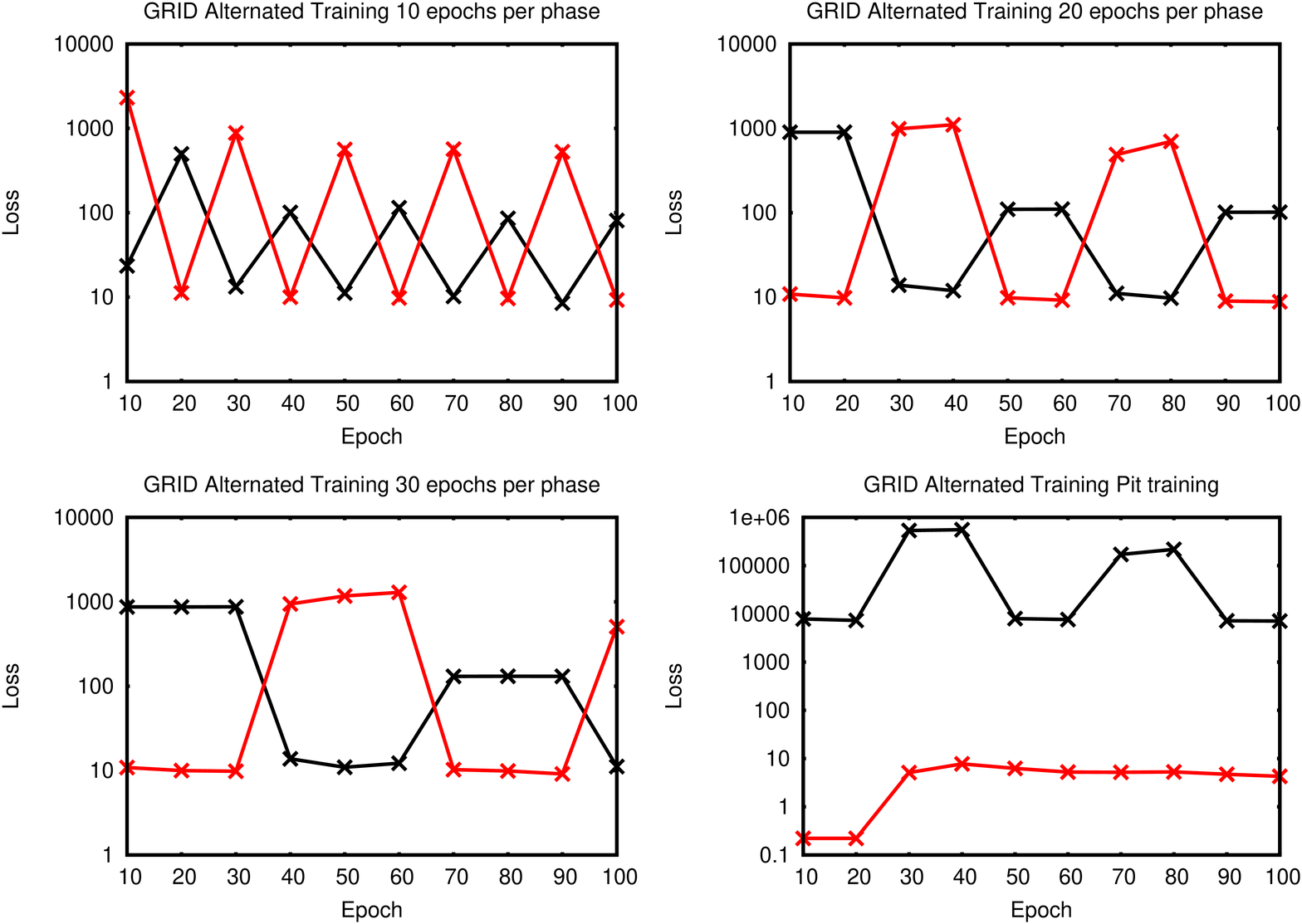}
\includegraphics[width=2.8in,keepaspectratio]{img/key.eps}
\caption{Trend of the two losses on the GRID validation set during training with \textit{alternated training}, by using different number of epochs per phase. \label{Fig:Res_altrenate}}
\vspace{-1em}
\end{figure}

In both datasets, the joint model significantly outperforms baselines with mixed-speech input. In particular, the \textit{alternated training} reaches better results in GRID while in TCD-TIMIT it is slightly outperformed by the \textit{alternated two full phases training} with \textit{weight freezing}. 
We evaluated the joint model also by substituting the loss $\mathcal{L}^{enh}$ by an MSE-based loss function trained by removing visual input information and by using permutation invariant training (PIT) optimization \cite{Kolbaek17}, a very effective audio-only technique. We reported the results in the last row of the Table~\ref{tab:Res_per} and in Figure~\ref{Fig:Res_altrenate} that show PIT performs worse than audio-visual counterparts. 

\begin{figure}[ht]
\centering
\includegraphics[width=2.8in,keepaspectratio]{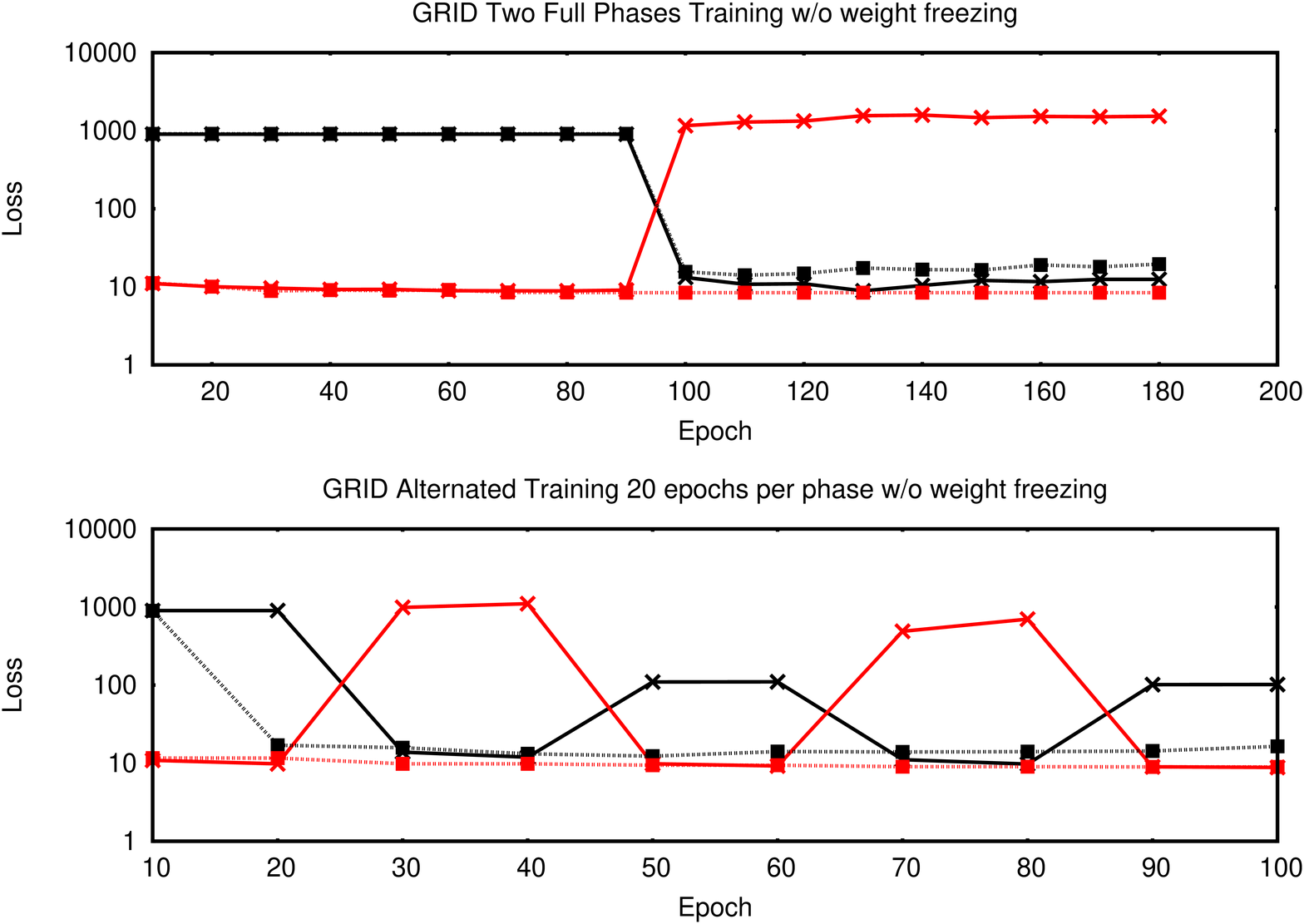}
\includegraphics[width=2.8in,keepaspectratio]{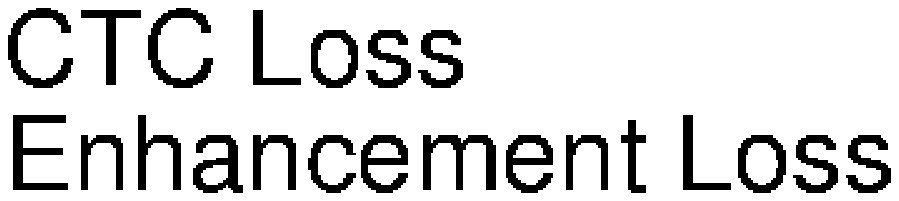}
\caption{Trend of the two losses on the GRID validation set during training with and without \textit{freezing weights} by using the \textit{alternated two full phases training} and \textit{alternate training}. \label{Fig:Res_two_step}}
\end{figure}

\begin{figure}[ht]
\centering
\includegraphics[width=2.8in,keepaspectratio]{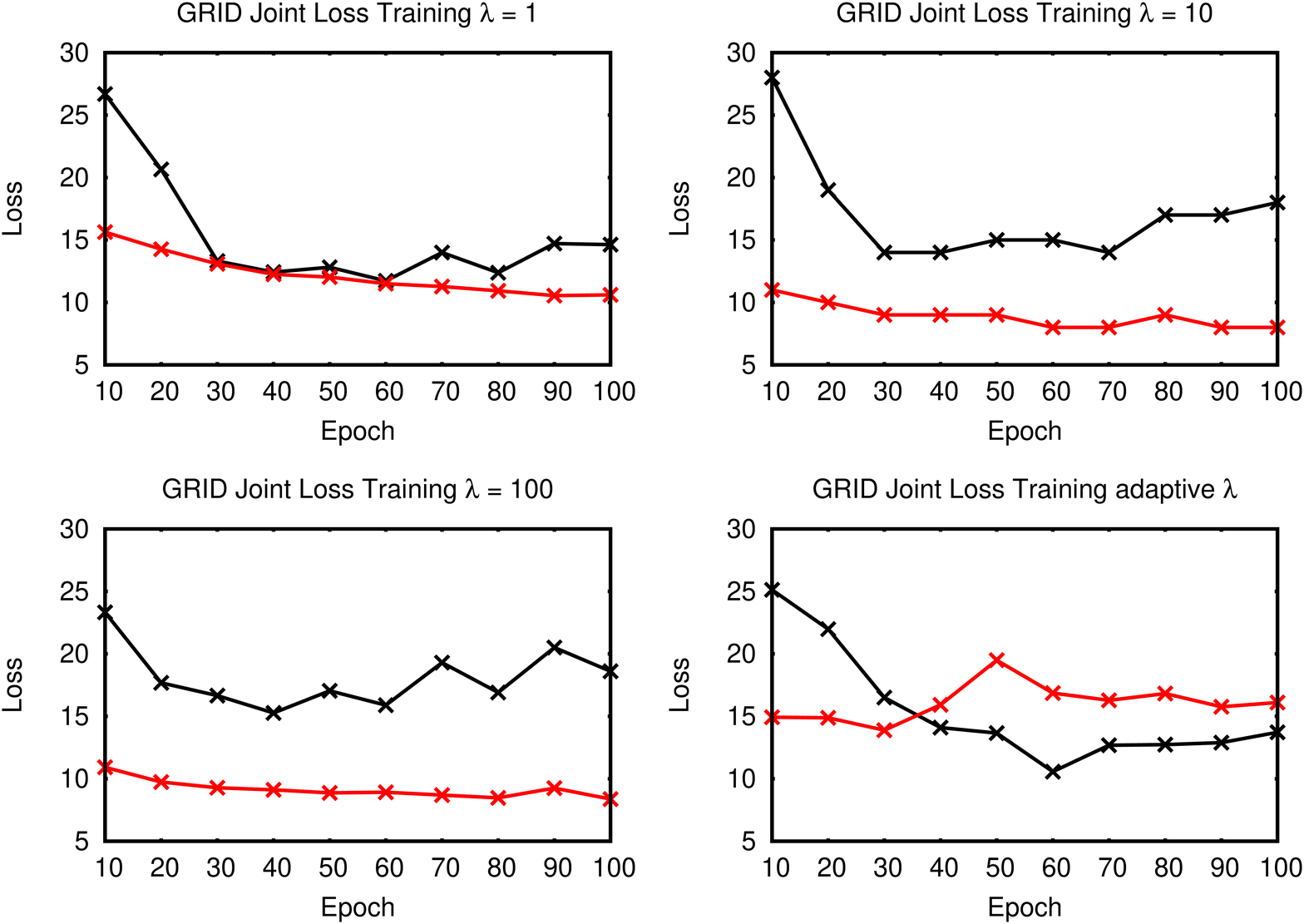}
\includegraphics[width=2.8in,keepaspectratio]{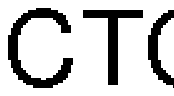}
\caption{Trend of the two losses on the GRID validation set during training by using the \textit{joint loss} with different $\lambda$ values. \label{Fig:Res_join}}
\vspace{-2em}
\end{figure}

\vspace{-1em}
\subsection{Result Analysis}
\vspace{-0.5em}
In this section, we analyze the trends of $\mathcal{L}^{enh}$ and $\mathcal{L}^{asr}$ during training, and in particular, we focus on their ratio. Due to space limitations, we only report, the loss curves computed on the GRID validation set, Figures~\ref{Fig:Res_two_step},~\ref{Fig:Res_join} and~\ref{Fig:Res_altrenate}. However, we observed an analogous behavior on TCD-TIMIT.

The first method that we analyze is the \textit{alternated two full phases training}. It first updates $\theta^{enh}$ parameters to minimize the $\mathcal{L}^{enh}$ loss, until it reaches a plateau in terms of speech enhancement on the validation set.

Figure~\ref{Fig:Res_two_step} shows that the alternated two full phases strategy from epoch 90, when the minimization of $\mathcal{L}^{asr}$ starts (and  involves both $\theta^{enh}$ and $\theta^{asr}$) the speech enhancement loss function $\mathcal{L}^{enh}$ remarkably diverges in few epochs.
This behavior suggests that the de-noised representation is not optimal to perform the phone recognition task, as observed in previous works \cite{wang2015joint,narayanan2015improving,chen2015speech}, although we did not expect to observe such a strong divergence. 
The $\mathcal{L}^{enh}$ and $\mathcal{L}^{asr}$ curves obtained by using \textit{alternated two full phases training} with \textit{weight freezing} unveil another effect of this issue. 
Here $\theta^{enh}$ parameters are forced to not change during the ASR training phase, and hence $\mathcal{L}^{enh}$ does not diverge but at the same time $\mathcal{L}^{asr}$ does not reach results as good as in the previous case. Figure~\ref{Fig:Res_two_step} shows a similar behaviour of alternate training when weight freezing is applied. 

The dramatic drop of the enhancement performance drove us to explore how the two losses evolve if they are trained together by using a \textit{joint loss} method. Figure~\ref{Fig:Res_join} shows the trends of $\mathcal{L}^{enh}$ and $\mathcal{L}^{asr}$ when using different fixed values of $\lambda$ and the adaptive $\lambda_{adapt}$ of equation~\ref{eq:lambda}. 
In this case while $\mathcal{L}^{enh}$ decreases, $\mathcal{L}^{asr}$ (after a certain point) tends to increase. 
For higher values of $\lambda$ the gap between the two loss functions increases, indeed $\mathcal{L}^{asr}$ tends to diverge rapidly during training.
The best result for $\mathcal{L}^{asr}$ is obtained using the adaptive $\lambda_{adapt}$ value (also for TCD-TIMIT). 
The enhancement capability continually grows as the epochs pass, while PER optimization has a substantial slowdown after 40 epochs. This deceleration coincides with the start of the faster decrease of the enhancement loss. The \textit{joint loss} training shows the interesting property of obtaining fair good results for both the metrics, but, in terms of ASR capability (that is the main goal of the model) the results turn out to be lower than the ones obtained with the some other training methods.

Figure~\ref{Fig:Res_altrenate} shows the trends of the two losses during \textit{alternated training}, with different number of epochs per phase. 
Even in this case the decrease of $\mathcal{L}^{asr}$ coincides with a large increase of the value of $\mathcal{L}^{enh}$ and vice-versa. 
Moreover, every repetition of the two phases leads to a smaller gap between the two loss functions.

\vspace{-1em}
\section{Conclusion}
\vspace{-0.5em}
In this paper we studied how audio-visual single channel speech enhancement can help speech recognition when several people are talking simultaneously.
The analysis unveils that jointly minimizing the speech enhancement loss and the CTC loss may not the best strategy to improve ASR. Then we explored the trends of the loss functions when the training strategy consists of an alternation of the speech enhancement and ASR training phases.
We observed that the loss function that was not considered for the training phase tends to diverge.
Finally, we found that the interaction between the two loss functions can be exploited in order to obtain better results. In particular, the \textit{alternated training} method shows that PER can be gradually reduced by wisely alternating the two training phases.


\bibliographystyle{IEEEbib}

\bibliography{main}
\end{document}